\documentclass{emulateapj}

\newcommand{\sref}[1]{Section~\ref{#1}}
\newcommand{\fref}[1]{Fig.~\ref{#1}}

\newcommand{\tref}[1]{Table~\ref{#1}}

\slugcomment{To be submitted to ApJ letters }

\shorttitle{Interpreting lensed supernova iPTF16geu}
\shortauthors{More et al.}

\usepackage{graphicx}
\usepackage{times}
\usepackage{amsmath}
\usepackage{natbib}
\usepackage{epstopdf}
\usepackage{color}

\begin{document}

\title{Interpreting the strongly lensed supernova iPTF16geu: Time
  delay predictions, microlensing, and lensing rates}
    
\author{
Anupreeta More\altaffilmark{1},
Sherry H. Suyu\altaffilmark{2,3,4},
Masamune Oguri\altaffilmark{1,5,6},
Surhud More\altaffilmark{1},
Chien-Hsiu Lee\altaffilmark{7}
}

\altaffiltext{1}{Kavli Institute for the Physics and Mathematics of the Universe (Kavli IPMU, WPI), University of Tokyo, Chiba 277-8583, Japan}
\altaffiltext{2}{Max-Planck-Institut f{\"u}r Astrophysik, Karl-Schwarzschild-Str.~1, 85748 Garching, Germany}
\altaffiltext{3}{Institute of Astronomy and Astrophysics, Academia Sinica,
P.O.~Box 23-141, Taipei 10617, Taiwan}
\altaffiltext{4}{Physik-Department, Technische Universit\"at M\"unchen, James-Franck-Stra\ss{}e~1, 85748 Garching, Germany}
\altaffiltext{5}{Department of Physics, University of Tokyo, 7-3-1 Hongo, Bunkyo-ku, Tokyo 113-0033, Japan} 
\altaffiltext{6}{Research Center for the Early Universe, University of Tokyo, Tokyo 113-0033, Japan}
\altaffiltext{7}{Subaru Telescope, National Astronomical Observatory of Japan, 650 North Aohoku Place, Hilo, HI 96720, USA}
\email{anupreeta.more@ipmu.jp}

\begin{abstract}
We present predictions for time delays between multiple images of the
gravitationally lensed supernova, iPTF16geu, which was recently
discovered from the intermediate Palomar Transient Factory (iPTF). As
the supernova is of Type Ia where the intrinsic luminosity is usually
well-known, accurately measured time delays of the multiple images could
provide tight constraints on the Hubble constant. According to our lens
mass models constrained by the {\it Hubble Space Telescope} F814W image,
we expect the maximum relative time delay to be less than a day, which
is consistent with the maximum of 100 hours reported by Goobar et al.
but places a stringent upper limit. Furthermore, the fluxes of most of
the supernova images depart from expected values suggesting that they
are affected by microlensing. The microlensing timescales are small
enough that they may pose significant problems to measure the time
delays reliably. Our lensing rate calculation indicates that the
occurrence of a lensed SN in iPTF is likely. However, the observed total
magnification of iPTF16geu is larger than expected, given its redshift.
This may be a further indication of ongoing microlensing in this system.
\end{abstract}


\section{Introduction}
Occurrence of strongly lensed supernovae (SNe) has long been predicted
\citep{Refsdal1964, Goobar2002, Oguri2003, Oguri2010b}, but they
had not been discovered until very recently.  \citet{Quimby2013, Quimby2014}
reported the discovery of a strongly lensed Type Ia supernova (SN Ia) PS1-10afx
with a total magnification of $\mu\sim 30$, although multiple images were not
resolved.  First resolved multiple images of a lensed SN were reported by
\citet{Kelly2015}: SN Refsdal, a core-collapse SN, that is strongly lensed into
multiple images by a foreground galaxy cluster.

Although many strongly lensed galaxies and quasars have already been
discovered, strongly lensed SNe have notable advantages
over traditional strong lenses, particularly if they are of Type
Ia. This is because of the standard candle nature of SNe Ia,
which allows us to measure the magnification factor directly. 
While measurements of the Hubble constant from time delays
\citep[e.g.,][]{SuyuEtal14, Wong2016, BonvinEtal16} need to overcome various degeneracies
including the mass-sheet degeneracy
\citep{FalcoEtal85,SchneiderSluse14},
the magnification factor provides important information on
the lens potential, which directly breaks the mass-sheet degeneracy
\citep{Kolatt1998} and the degeneracy in the lens potential and the
Hubble constant from time-delay measurements \citep{Oguri2003b}. 
Indeed several lensed SNe Ia behind massive clusters, although not
multiply imaged, have been used to constrain mass 
distributions of foreground clusters 
\citep{Riehm2011,Patel2014,Rodney2015}.

Recently, \citet[][hereafter, G16]{Goobar2016} reported the discovery of
a new gravitationally lensed Type Ia supernova ( SN Ia) iPTF16geu from
intermediate Palomar Transient Factory (iPTF). In this letter, we
present time-delay predictions and interpret SN magnifications in light
of microlensing. This letter is organised as follows. We introduce
iPTF16geu in \sref{sec:iptf_sn} and describe our mass modelling method
in \sref{sec:mod}.  We present predictions for magnifications and time
delays and discuss the role of microlensing in \sref{sec:res}. We 
calculate the expected frequency of lensed SNe Ia in \sref{sec:n_sne}
and give our conclusions in \sref{sec:conc}. 

\section{iPTF16geu}
\label{sec:iptf_sn}
\begin{figure*}
\plotone{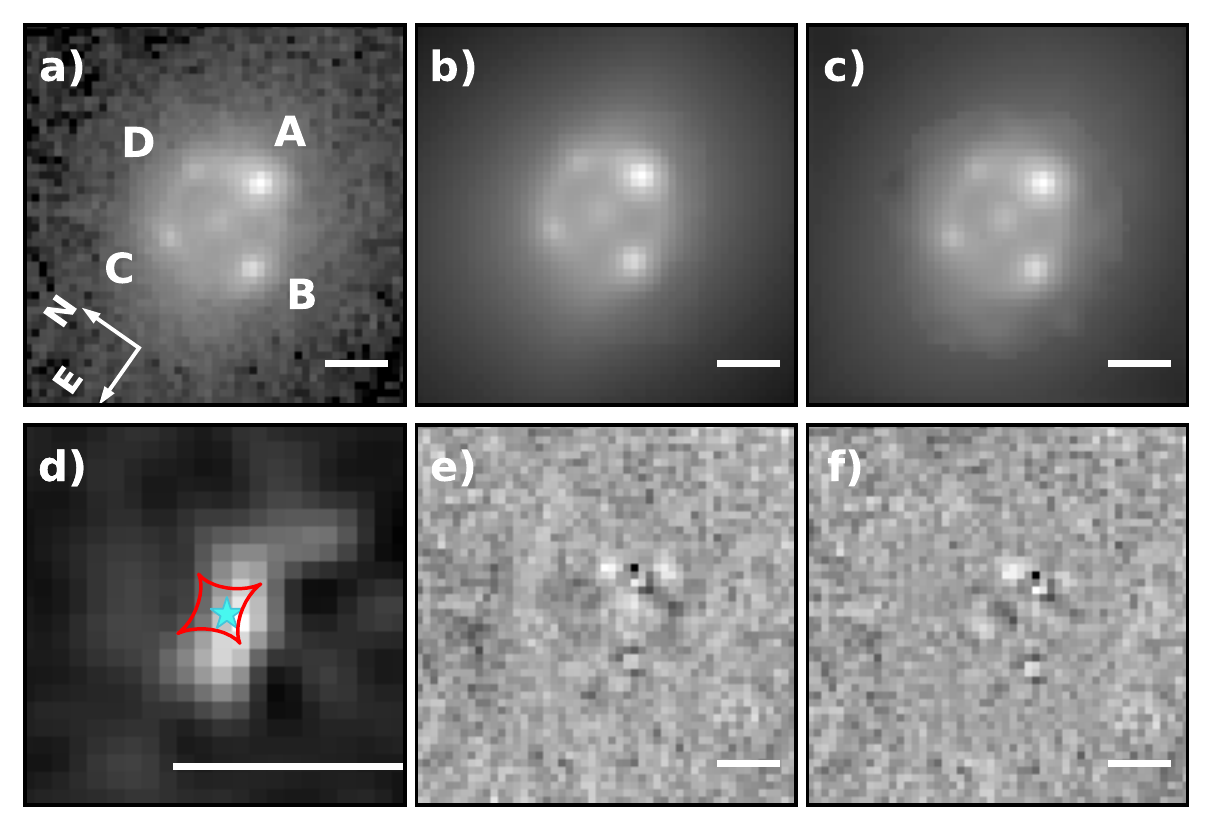}
\caption{ a) HST image (F814W) of iPTF16geu (Oct 28, 2016). Lens
mass models from b) {\sc glafic} and c) {\sc Glee} and normalised
residual images (e and f) in the bottom row, respectively.  d)
The reconstructed surface brightness distribution of the SN host galaxy
from the most probable lens model of {\sc Glee}. Caustics (red curves)
and the location of SN (blue star) are also shown. Images are 2\arcsec
on the side, except for panel d which is 0.5\arcsec and all scale bars correspond to
0.3\arcsec. 
\label{fig:mod}}
\end{figure*}

The  SN Ia of iPTF16geu has a redshift $z_{\rm SN}$ = 0.409 and is
magnified by a factor of $\mu \sim 56$ by an intervening galaxy at $z_{\rm l}
= 0.216$ (G16). After spectroscopic confirmation, several follow-up
programs were triggered to resolve the multiple images, measure light
curves and time delays.  Among the ground-based follow-up, data
taken with OSIRIS on Keck with adaptive optics on Oct 13, 2016 yielded
an image quality with FWHM=0\farcs07 in $H$ band establishing the
presence of multiple lensed images of the SN. However, only two brighter
SN images were visible because light from the host galaxy dominated the
emission at near-infrared (NIR) wavelengths. Subsequent high-resolution
images taken by the {\it Hubble Space Telescope} ({\it HST}) in the
optical clearly revealed four SN images (DD 14862, PI: Goobar). In
this letter, we use the \textit{HST} image taken on Oct 28, 2016, since
it is deeper than images taken at previous epochs. We choose the F814W
band which shows all SN images clearly (labelled A-D in descending
order of their fluxes, see \fref{fig:mod}).

\section{Lens mass modelling}
\label{sec:mod}
We model iPTF16geu with two different mass modelling software: {\sc
glafic} \citep{Oguri2010} and {\sc Glee} \citep{Suyu2010,SuyuEtal12a}.
This work is done independently by different coauthors, providing cross
validation of our model results and predictions.  Our cosmology is set
to $\Omega_m=0.32$, $\Omega_\Lambda=0.68$, and $h=0.72$. The
corresponding time-delay distance for the lens system is $D_{\Delta
t}=1920$\,Mpc. 

\begin{figure}
\begin{center}
\epsscale{1.2}
\plotone{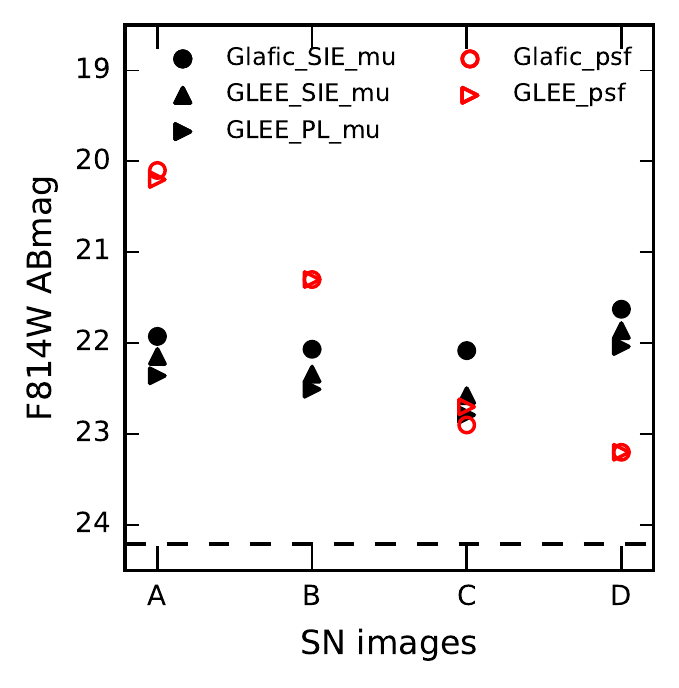}
\caption{Fluxes of SN images A, B, C and D. Expected fluxes after
scaling the intrinsic SN flux (24.21 ABmag, dashed line) by the lens-model
magnification ($\mu$) factors (filled symbols) are compared with PSF
model fluxes fit to the {\it HST} image. Relative magnifications are
more robust than the absolute values across different models.  Fluxes of
most of the images depart from predictions. Image A is the most
magnified and image D appears to be suppressed (see text in
\sref{sec:res} for further discussion).  
\label{fig:snmag}}
\end{center}
\end{figure}

\begin{table*}
\begin{center}
\caption{Model parameters and predictions.\label{tab:mod}}
\begin{tabular}{lcrrrrccccc}
\tableline
\tableline
 \multicolumn{6}{c}{Lens Model}   & & \multicolumn{4}{c} {$\kappa,\gamma$}\\  
\tableline
Model Profile & $\theta_{\rm E}$(\arcsec)  & $q_{\rm m}$  &
$\varphi_e$ & $\gamma'$ & $\gamma_{\rm ext}, \varphi_{\rm ext}$ & \phantom{space}& A & B & C & D  \\
\tableline
{\sc glafic} SIE & $0.29\pm0.01$ & $0.83\pm0.01$ & $65\pm1$ & $\equiv2.0$ & -- 
&& 0.56,0.56 & 0.43,0.43 & 0.57,0.56 & 0.46,0.45 \\
{\sc Glee} SIE & $0.294\pm0.002$ & $0.77^{+0.03}_{-0.02}$ & $66\pm1$ & $\equiv2.0$ & -- 
&& 0.60,0.60 & 0.40,0.40 & 0.62,0.62 & 0.43,0.43 \\
{\sc Glee} PL  & $0.30\pm0.01$ & $0.73\pm0.04$ & $66\pm1$ & $2.1\pm0.1$ &  --
&& 0.56,0.66 & 0.35,0.44 & 0.58,0.68 & 0.38,0.48 \\
{\sc Glee} PL+$\gamma_{\rm ext}$ & $0.30\pm0.01$  & $0.66^{+0.08}_{-0.04}$ & $68^{+4}_{-2}$ & $2.1\pm0.1$ &  $0.02^{+0.03}_{-0.01}, 79^{+8}_{-14}$ 
&& 0.63,0.61 & 0.36,0.44 & 0.64,0.64 & 0.40,0.47 \\
\tableline
\tableline
\end{tabular} 
\begin{tabular}{lccccccccc}
\tableline
\tableline 
\multicolumn{10}{c}{Model Predictions} \\
\tableline
Model & \multicolumn{4}{c} {Magnification factors} & & \multicolumn{4}{c} {$\Delta t$ (days)} \\
Profile  & A & B & C & D & \phantom{space} & A & B & C & D \\
\tableline
{\sc glafic} SIE & $-8.2_{-0.5}^{+0.4}$ & $7.2_{-0.2}^{+0.2}$ & $-7.1_{-0.3}^{+0.3}$ & $10.8_{-0.4}^{+0.4}$ & &
$0.40_{-0.02}^{+0.02}$ & $\equiv0$    &  $0.47_{-0.02}^{+0.01}$ & $0.25_{-0.01}^{+0.01}$ \\
{\sc Glee} SIE & $-6.7^{+1.2}_{-1.0}$ & $5.6^{+0.6}_{-0.6}$ & $-4.5^{+0.6}_{-0.6}$ & $8.7^{+1.1}_{-1.3}$ & & 
$0.52^{+0.08}_{-0.05}$ & $ \equiv0 $ & $0.65^{+0.07}_{-0.07}$ & $0.35^{+0.05}_{-0.05}$   \\
{\sc Glee} PL & $ -5.5^{+0.9}_{-1.5} $ & $ 4.8^{+0.9}_{-0.6}$ & $ -3.7^{+0.5}_{-0.9}$ & $ 7.4^{+1.6}_{-0.9} $ & & 
$ 0.56^{+0.06}_{-0.06} $   &  $ \equiv0 $  &  $ 0.70^{+0.06}_{-0.07} $  &  $ 0.37^{+0.03}_{-0.04} $  \\
{\sc Glee} PL+$\gamma_{\rm ext}$ & $-5.2^{+1.7}_{-1.9}$ & $4.7^{+1.3}_{-1.2}$ & $-3.6^{+1.2}_{-1.3}$ & $7.4^{+1.9}_{-2.0}$ & &
$0.6\pm0.1$ &  $\equiv 0$ & $0.7\pm0.1$ & $0.4\pm0.1$\\
\tableline
\tableline
 \end{tabular}
  
\tablecomments{$\theta_{\rm E}$ is the Einstein radius.  $q_{\rm m}$ is
the axis ratio of the lens mass.  The PAs ($\varphi_e$ and $\varphi_{\rm
ext}$) are in degrees measured East of North. A shear angle of
$\varphi_{\rm ext}=0$ corresponds to shearing of the lens system along
the north-south direction. The most-probable convergence ($\kappa$) and shear
($\gamma$) values are given at the location of each SN image. Negative
magnification ($\mu$) means opposite parity and $\Delta t$ is time delay
relative to image B.  }

\end{center}
\end{table*}

The four SN images in iPTF16geu are almost equidistant from each other
in a cross-like configuration 
where 
we expect the multiple images to be magnified
by similar factors unlike what we see in iPTF16geu.  Since the fluxes
can be affected due to effects such as microlensing and time delay, we
model each supernova image as a point spread function with a free
amplitude in the image plane.  Additional data constraints come from the
extended host galaxy which is lensed into almost an Einstein ring. Both
the software model the light of foreground lens galaxy with a Sersic
profile \citep{Sersic68}, but differ in their assumptions about the mass
profile of the lens and the model for the SN host galaxy.
 
\subsection{Parametric source model} \label{sec:gla}
We fit the arbitrary SN fluxes simultaneously as we fit the SN positions
and the lensed host (Sersic) with lens mass model using {\sc glafic}. The lens
mass distribution is modelled as a singular isothermal ellipsoid (SIE). We
imposed the following constraints on the lens parameters. The
centroid, axis ratio ($q_{\rm m}$) and position angle (PA, $\varphi_e$) is
assumed to be the same for the mass density and light profiles. External
shear ($\gamma_{\rm ext}$) is often degenerate with the ellipticity of
the mass distribution. Hence, we did not include any external shear, and
were able to find a good model fit (see middle column in
\fref{fig:mod}). We used a custom {\sc emcee} \citep{ForemanMackey2013}
wrapper around {\sc glafic} to sample the posterior distribution of our
models using Markov chain Monte Carlo (MCMC) approach.

\subsection{Pixellated source model} \label{sec:glee} 
With {\sc Glee}, we fit the SN images on the image plane simultaneously
with its host galaxy surface brightness that is modeled on a grid of
pixels on the source plane \citep{SuyuEtal06}.  We use a power-law mass
distribution for the lensing galaxy \citep[e.g.,][]{Barkana98}, with six
parameters: centroid ($\theta_{\rm m1}$, $\theta_{\rm m2}$), $q_{\rm
m}$, $\varphi_{\rm e}$, Einstein radius $\theta_{\rm E}$ and radial mass
profile slope $\gamma'$ corresponding to the three-dimensional mass
density $\rho \propto r^{-\gamma'}$. We also test SIE model for
comparison with results of {\sc glafic}.  We further consider a lens
model that includes an external shear component, with the shear
magnitude ($\gamma_{\rm ext}$) and angle ($\varphi_{\rm ext}$) as two
additional parameters.  The lensed arcs of the SN host galaxy and the
fitted SN image positions provide constraints on the lens mass
parameters.  The SN and lens mass/light parameters have uniform priors.
We sample all the model parameters using either {\sc emcee} or the MCMC
method described in \citet{DunkleyEtal05}.

\section{Results and Discussion}
\label{sec:res}
In \fref{fig:mod}, we show images of our most probable models (top row -
panels b and c) and the corresponding residuals (model subtracted from the
data, normalized by the estimated pixel uncertainties, bottom row -
panels e and f). The pixellated reconstruction of the SN host galaxy on
the source plane from {\sc Glee} is also shown (panel d) with the
location of SN (star). Parts of the host galaxy and the SN lie within
the astroid caustic (red curve) and are thus quadruply imaged, whereas
other parts are doubly imaged. 

We also present our modelling results from both {\sc glafic} and {\sc
Glee} in \tref{tab:mod} and they agree reasonably well.  The median of
the posterior distributions and their 68\% confidence levels for the
lens parameters are given. The slope of the density profile is
consistent with isothermal ($\gamma'=2.0$) within the uncertainties. The
power law models with and without $\gamma_{\rm ext}$ are the same within
the uncertainties suggesting that the role of $\gamma_{\rm ext}$ is not
significant. We give the convergence ($\kappa$) which is the
surface mass density in the units of critical surface mass density and
shear values at the location of SN images A, B, C and D in
\tref{tab:mod}. We also give predictions for magnifications and relative
time delays. Negative magnifications imply opposite parity of the
images and that they correspond to saddle points in the time-delay
surface \citep[see e.g.,][]{Blandford1986}. As expected, the predicted
relative magnifications are comparable (within a factor of 2) for all SN
images.  Light from image B arrives first, followed by images D, A and C
consistently for all our models. Images A and C are saddle images
(negative parity) and images B and D are minima (positive parity). We
find that the time delays (relative to image B) are within a day, a more
stringent limit than the 100-hour range predicted by G16. 

Now, we use the standard candle nature of SN Ia to our advantage in
understanding expected and observed SN fluxes.  We take the best-fit
Sloan Digital Sky Survey (SDSS) i-band model light curve from G16
(corresponding to F814W). After accounting for time dilation and dust
extinction of E(B-V)=0.31 assuming Rv=3.1, we calculate the un-lensed SN
flux expected on Oct 28, 2016 (35 days from maximum) to be 24.21 ABmag.
In \fref{fig:snmag}, we show the un-lensed SN flux scaled up by
magnification factors from our models (filled symbols). For comparison,
we also plot the observed PSF fluxes of the SN images (open symbols,
based on fits by the two modeling software).  It is interesting to note
that image D which appears the faintest in the data is predicted to have
the largest magnification factor.  Image C is the least magnified of all
and most consistent with expected magnification.  Both images A and B
are expected to be fainter than D in the absence of any external factors
affecting the fluxes but are found to be greatly magnified.  

Since the relative time delays are less than a day from our models,
differences in the observed fluxes are unlikely to be due to time
delays. While differential dust extinction could be a possible cause of
anomalies in fluxes, it is unlikely to produce such large differences,
given the almost symmetrical distribution of SN images around the lens
which is an early-type galaxy \citep[e.g.][also, G16 suggest that
extinction is not signficant]{Falco1999,Eliasdottir2006}. The most
likely explanation for highly anomalous fluxes is microlensing due to
stars in the foreground lens galaxy\footnote{We do not consider lensing
by subhalo or milli-lensing explicitly because it essentially has the
same consequences qualitatively as microlensing in single epoch images,
but arising due to a more massive subhalo instead of stars. With
superior resolution and multi-epoch data and, it may be possible to
disentangle between millilensing (static) and microlensing (time
varying) effects.  We restrict our discussion to microlensing for
iPTF16geu which we believe has a higher optical depth, but we
acknowledge the possible presence of milli-lensing.}. 

Saddle images are more susceptible to show microlensing
(de-)magnification in their fluxes whereas minimum images typically have
enhanced fluxes over the macro-magnification (i.e. magnification from a
smooth mass model e.g., \citealt{Kochanek2006b}). Indeed, we find
that image A, located at a saddle point, is brighter than our smooth
mass model predictions by nearly two  magnitudes for the epoch presented
here.  Images B and D, which are at minima, are found to be magnified
and suppressed, respectively, by nearly a magnitude each.  Image C does
not seem to be affected by microlensing. However, we need to analyse
multi-epoch data in order to be certain about the extent to which each
of the images are affected by microlensing. This work is left to a
future paper.
 
\citet{Dobler2006} calculated the optical depths for microlensing of
lensed  SN Ia and found a high probability for lensed SNe to be affected
by microlensing. Using typical stellar mass fractions measured from
the lens galaxies of the Sloan ACS Survey \citep{Bolton2004}, they found
that 25\% of lensed SN Ia showed differences of more than 1 magnitude.
The stellar mass fraction ($<\theta_{\rm E}$) for the lens in iPTF16geu
is $\sim0.9$ derived using stellar mass - velocity dispersion
(M$\ast-\sigma$) relation \citep[][and references therein]{Zahid2016}.
This suggests a high optical depth although there is a large uncertainty
in the mass fraction due to the scatter (about a factor of 2) in the
M$\ast-\sigma$ relation. For a typical SN photosphere size
($10^{15}$~cm) and lens velocity dispersion ($\sim 150$~km~s$^{-1}$), we
calculate caustic crossing time to be $\sim 2$~years
\citep[e.g.,][]{Treyer2004}.  On the other hand, SN photosphere
velocities of $15000$~km~s$^{-1}$ imply that microlensing effects could
be visible on shorter time scales corresponding to the light curve.
Microlensing can affect SN Ia light curves on different time scales and
produce different qualitative signatures  (e.g., large magnitude
offsets, introduce non-intrinsic peaks, change the decay rate) thus
making it difficult to measure time delays with accuracy better than a
few days \citep{Dobler2006}. 

\section{Frequency of lensed  SN Ia}
\label{sec:n_sne}
G16 used the SNOC Monte-Carlo package
\citep{Goobar2002b} to calculate the expected number of lensed SNe Ia
with high magnification and found that it appears to be too low to
explain the discovery of iPTF16geu. Based on this comparison, they
argued that lensing by sub-kpc structures may have been greatly
underestimated. Here we present an independent comparison of the
expected number and property of strongly lensed SN Ia with iPTF16geu.  

Our estimate of the expected lensing rate is based on Monte-Carlo
simulations presented in \citet{Oguri2010b}, in which realistic
population of lensing galaxies and the source population has been
considered, and various selection biases such as magnification bias 
and K-correction have been properly taken into account. Mass
distributions of lensing galaxies are assumed to follow the singular
isothermal ellipsoid with an external shear. Here the velocity
function of galaxies of all types directly measured in SDSS
\citep{Bernardi2010} is used for the abundance of lensing
galaxies. However, calculations of \citet{Oguri2010b} assumed that 
multiple images be resolved in surveys, which was not the case for
iPTF. The poor spatial resolution of iPTF makes most of multiply
imaged SNe blended. In order to compute lensing rates in such a
situation, we use the total magnification, rather than magnifications
of individual images used in \citet{Oguri2010b}, to compute the
magnification bias. We also impose no lower limit on the image
separation \citep[see also][]{Quimby2014}.

In order to compute the expected total number of lensed SNe Ia, we
need to know the total survey volume of iPTF. According to
G16, the total monitoring time and the average solid
angle of iPTF and PTF (precursor survey to iPTF) translates into the
total survey volume of $\sim 5000$~deg$^2$year, which we adopt in the
following calculations. The detection limit is $R\sim 21$, but in
order to observe light curves well we assume that peak magnitudes of
SNe be one magnitude brighter than the limiting magnitude in order to
be detected and studied in iPTF, i.e., the limiting magnitude of the
peak SN brightness of $R=20$.

\begin{figure}
\epsscale{1.2}
\plotone{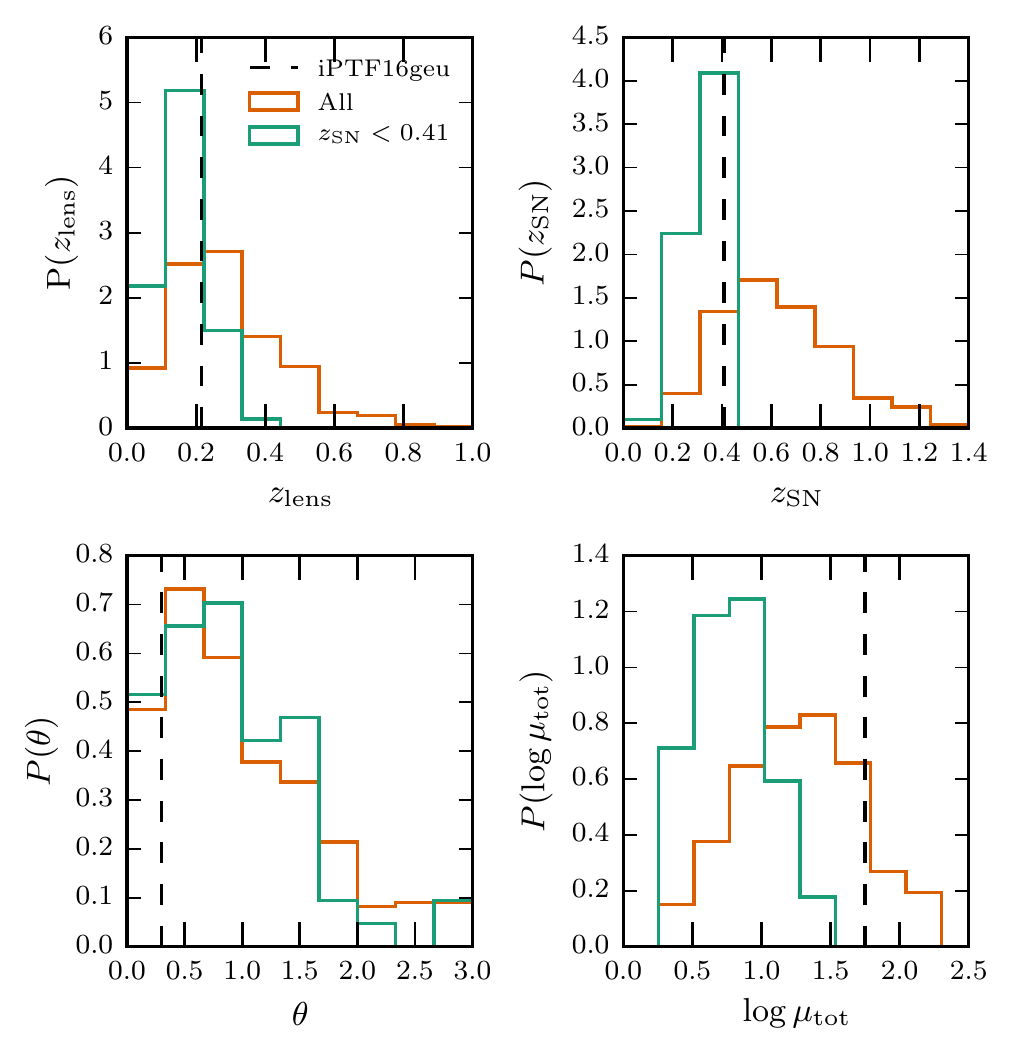}
\caption{Expected distributions of the lens redshift $z_{\rm l}$ ({\it upper
    left}), source (SN) redshift $z_{\rm SN}$ ({\it upper
    right}), image separation $\theta$
  ({\it lower left}), and total magnification $\mu_{\rm tot}$ ({\it
    lower right}), in iPTF and PTF, which are computed using the
  method described in \citet{Oguri2010b} with some modifications to
  match the selection function of PTF/iPTF. 
    \label{fig:dist}} 
\end{figure}

We find that the expected number of lensed SNe Ia in iPTF and PTF calculated
using the setup above is $0.9$, which would be consistent with the discovery of
iPTF16geu. The probability distribution for the expected redshifts, image
separation, and magnifications are shown using orange histograms and compared to
iPTF16geu (vertical dashed line) in \fref{fig:dist}. Because of the large effect of the
magnification bias, our calculation also predicts a high fraction of quadruple
lenses, $\sim 65$\%, which is consistent with iPTF16geu being a quadruple
lens. However, the efficiency of the spectroscopic typing of iPTF events goes
down at $z\gtrsim0.41$ (Goobar, priv. comm.). The expected number of lensed SNe
Ia with $z_{\rm SN}<0.41$ is $0.16$ (see green histograms for distribution of
properties), making it less likely to be discovered. The observed magnification
of iPTF16geu in this case also appears inconsistent given the distribution. This
is in agreement with G16, who require extreme assumptions about the fraction of
compact objects in halos. This may be further evidence for the presence
of microlensing.

\section{Conclusion}
\label{sec:conc}
We have presented lens modelling results for the recently discovered
gravitationally lensed  SN Ia, iPTF16geu (G16). Our mass modeling
predicts flux ratios within a factor of 2 across the four lensed images
of the SN. However, the brightest images A and B are nearly 15 and 5
times brighter, respectively, than the fainter pair of images (C and D)
as measured from the HST F814W image. Differential extinction may have
very little to no contribution, as noted in G16, especially at NIR. And,
yet the high contrast in flux ratios appears similar at other optical
and NIR wavelengths (see G16). 

Our interpretation is that most of the SN images are affected by
microlensing. We derived the unlensed SN flux for the epoch of Oct
28 using the best-fit model light curve (G16). Multiplying the
unlensed SN flux by our lens model magnification factors suggests that,
in addition to the macro-magnification, fluxes of images A and B are
further magnified by more than a magnitude and image D is suppressed by
nearly a magnitude due to microlensing. While image C does not seem to
be affected by microlensing from the current analysis, we need to
analyse multi-epoch data to understand how each of the SN
images are affected by microlensing.

We predict relative time delays of the order of less than a day, a consistent
but more stringent upper limit than the hundred hours predicted in G16. Accurate
measurements of the time delays will require observations with high cadence
($\sim$ a few hours apart) and preferably around characteristic features in the
light curves. However, small time delays predicted in our mass modeling and
probable microlensing effects suggest that accurate measurements of time delays
may be quite challenging \citep[e.g.,][]{Dobler2006}.

Lastly, based on our detailed calculations of lensing rates, the
expected average number of lensed  SN Ia from PTF/iPTF is $0.9$. However
the spectroscopic followup and typing efficiency implies a restriction
of $z_{SN}<0.41$ (G16 and A.~Goobar priv.~comm.) which reduces this
expected number to $0.16$. This also implies that the high observed
magnification of iPTF16geu is quite unlikely, and hints at a possible
role of microlensing.

\acknowledgments
We thank Ariel Goobar, Sergey Blinnikov, Alexey Tolstov and Matteo
Barnab{\`e} for helpful discussions and the referee for useful
suggestions. This work is supported by World Premier International
Research Center Initiative (WPI Initiative), MEXT, Japan. MO is
supported by JSPS KAKENHI Grant Number 26800093 and 15H05892. SHS
gratefully acknowledges support from the Max Planck Society through the
Max Planck Research Group. Based on observations made with the NASA/ESA
Hubble Space Telescope, obtained from the data archive at the Space
Telescope Science Institute. STScI is operated by the Association of
Universities for Research in Astronomy, Inc. under NASA contract NAS
5-26555.  

Facilities: \facility{HST(WFC3)}


\end{document}